\documentstyle[twocolumn,aps,prb,epsfig]{revtex}
\begin{document}
\newlength{\figwidth}
\setlength{\figwidth}{3.25in}

\twocolumn[\hsize\textwidth\columnwidth\hsize\csname %
@twocolumnfalse\endcsname

\draft
\widetext

\title{Monte Carlo Study of the S=1/2 and S=1 Heisenberg
Antiferromagnet on a Spatially Anisotropic Square Lattice}
\author{Y. J. Kim and R. J. Birgeneau}
\address{Department of Physics, Massachusetts 
Institute of Technology, Cambridge, Massachusetts 02139}
\date{\today}
\maketitle

\begin{abstract} We present a quantum Monte Carlo study of a Heisenberg
antiferromagnet on a spatially anisotropic square lattice, where the
coupling strength in the x-direction ($J_x$) is different from that in the
y-direction ($J_y$). By varying the anisotropy $\alpha$ from 0 to 1, we
interpolate between the one-dimensional chain and the two-dimensional
isotropic square lattice. Both $S=1/2$ and $S=1$ systems are considered
separately in order to facilitate comparison. The temperature dependence
of the uniform susceptibility and the spin-spin correlation length are
computed down to very low temperatures for various values of $\alpha$.  
For $S=1$, the existence of a quantum critical point at
$\alpha^{S=1}_c=0.040(5)$ as well as the scaling of the spin gap is
confirmed. Universal quantities predicted from the ${\cal O}(3)$ nonlinear
$\sigma$ model agree with our results at $\alpha=0.04$ without any
adjustable parameters.  On the other hand, the $S=1/2$ results are
consistent with $\alpha^{S=1/2}_c=0$, as discussed by a number of previous
theoretical studies. Experimental implications for $S=1/2$ compounds such
as Sr$_2$CuO$_3$ are also discussed. 
\end{abstract}

\phantom{.}
]

\narrowtext


\section{Introduction}

Low-dimensional quantum magnetism is currently one of the most intensively
studied fields in condensed matter physics.  Through the synergistic
efforts of theory, numerical simulation, and experiment, our understanding
of the static and dynamic behavior of low dimensional quantum magnets has
grown tremendously.  The study of the one-dimensional (1D) quantum
Heisenberg antiferromagnet (QHA) has a long history dating back to the
exact solution found by Bethe \cite{Bethe31} for the $S=1/2$ nearest
neighbor (NN) chain.  He found the ground state eigenfunction for this
system and showed that there is no long-range order at $T=0$, unlike its
classical counterpart.  The Bethe Ansatz solution, however, is peculiar to
the $S=1/2$ chain and cannot be readily generalized. Recent theoretical
advances in quantum magnetism have come primarily from applications of
quantum field theory techniques. The low-energy, long-wavelength behavior
of quantum antiferromagnets can be mapped onto an effective relativistic
field theory.

In his pioneering work based on the large-spin, semi-classical mapping of
the 1DQHA to the ${\cal O}(3)$ quantum nonlinear $\sigma$ model
(QNL$\sigma$M) in (1+1) dimensions, \cite{Haldane83a} Haldane conjectured
that all half-odd-integer spin chains should behave qualitatively like a
$S=1/2$ chain, while for integer spin chains the zero-temperature spin
correlations should decay exponentially with distance due to the presence
of the so-called Haldane gap, $\Delta$.  In two-dimensions, Chakravarty,
Halperin, and Nelson \cite{Chakravarty89} have mapped the long-wavelength,
low-temperature behavior of the two-dimensional (2D) QHA onto the (2+1)
dimensional QNL$\sigma$M. They obtain a phase diagram for the system with
three regimes: quantum disordered (QD), quantum critical (QC), and
renormalized classical (RC). In the QD regime, the spin correlation length
$\xi$ remains finite even at $T=0$, as in the case of the $S=1$ spin
chain.  In the QC regime, temperature is the only relevant energy scale,
and thus $\xi$ diverges like $\sim T^{-1}$ as $T \rightarrow 0$: $T=0$ is
the quantum critical point. In the RC regime, the correlation length
diverges exponentially [$\sim \exp(1/T)$], so that the ground state has
long-range order.

Due to the peculiar role played by the spin quantum number in the ground
state properties, the 1DQHA has been much studied since Haldane's
conjecture. The 2DQHA has also attracted considerable attention both for
its intrinsic interest and also because the magnetism in the parent
compounds of the high temperature superconductors is well described by the
$S=1/2$ 2DQHA. Recently, attention has been given to the physics of
magnetism in intermediate dimensions; namely, systems which exhibit a
crossover from one to two dimensions. One can think of two different
approaches to explore this problem. One method is to begin with spin
ladders, which are obtained by coupling a small number of spin chains, and
then to increase the number of chains in the spin ladder until 2D physics
is obtained. Alternatively, one can begin with an infinite number of
decoupled spin chains, and then increase the coupling between the spin
chains gradually until the interchain coupling becomes comparable to the
intrachain coupling; in other words, one studies a spatially anisotropic
square lattice quantum Heisenberg antiferromagnet (SASLQHA). Spin ladders
have been studied much recently, and have revealed the initially
surprising result that the ground state properties of the $S=1/2$ spin
ladder depend on the number of chains in the ladder, \cite{Dagotto96}
analogous to the 1DQHA. Specifically, when even numbers of chains (or
legs) are coupled to form a ladder (even-$n$), they show the same
universal behavior as integer spin chains, while odd-$n$ $S=1/2$ ladders
behave essentially like a single $S=1/2$ chain at low temperatures and
long wavelengths.

Our study focuses on the SASLQHA, in which the dimensional parameter
(interchain coupling) can be varied {\it continuously}. The Hamiltonian of
the SASLQHA is essentially that of parallel spin chains forming a square
lattice with interchain coupling $J_x$ ($J_x > 0$):
\begin{equation}
{\cal H} = J_y \sum_{i,j}{\bf S}_{i,j} \cdot {\bf S}_{i+1,j}
+\sum_{i,j} J_x {\bf S}_{i,j} \cdot {\bf S}_{i,j+1}.
\label{hamil}
\end{equation}
Here $i$ and $j$ label lattice sites along the y-direction and the
x-direction, respectively. We will set $J_y=J$ and use $\alpha \equiv J_x
/ J_y$ as an anisotropy parameter in this paper.  We use units in which
$\hbar=k_B=g \mu_B=1$, and set the lattice constant $a=1$ along both the x
and y directions. By varying $0 \leq \alpha \leq 1$, the Hamiltonian Eq.\
(\ref{hamil}) can interpolate between a single spin chain and an isotropic
square lattice. The $S=1/2$ case has drawn special attention recently, due
to its conjectured relevance to the physics of stripe structures in high
temperature superconductors.  \cite{CastroNeto96,vanDuin98}

One of the most interesting questions in the physics of the SASLQHA is the
nature of the ground state. Since the ground state of the 1DQHA
($\alpha=0$) is disordered while the QHA on an isotropic square lattice
($\alpha=1$) has a N\'eel ordered ground state, there should be an
order-disorder quantum transition for a critical value of
$\alpha=\alpha_c$.  For $S=1$ this seems intuitively clear. As $\alpha$ is
increased from zero, the Haldane gap will decrease smoothly, vanishing at
a non-zero $\alpha_c^{S=1}$. However, it is not at all obvious what the
behavior should be for $S=1/2$.  Although it is widely believed that
$\alpha_c^{S=1/2}=0$, there are a few studies claiming
otherwise. \cite{CastroNeto96,Parola93} Sakai
and Takahashi \cite{Sakai89} first considered the Hamiltonian Eq.\
(\ref{hamil})  by treating the interchain coupling in the small $\alpha$
limit via mean field theory. They also estimated that $\alpha_c^{S=1/2}
\approx 0$ and $\alpha_c^{S=1} \approx 0.025$ by calculating the
susceptibility via numerical Lanzcos diagonalization. Azzouz and coworkers
obtained essentially the same results from their field theoretical study,
albeit with smaller $\alpha_c^{S=1} \approx 0.00186$.
\cite{Azzouz93a,Azzouz93b} The renormalization-group argument by Affleck
and coworkers \cite{Affleck94,Affleck96} suggests that the answer is not
universal but depends on the microscopic details of the model. For the
nearest-neighbor model [Eq.\ (\ref{hamil})], $\alpha_c^{S=1/2}=0$ was
shown by zero temperature series expansions.

In this paper we show that our quantum Monte Carlo study of the $S=1/2$
SASLQHA gives results which are consistent with the claim that N\'eel
order sets in for infinitesimal $\alpha$ for $S=1/2$.
\cite{Sakai89,Azzouz93b,Affleck94,Miyazaki95,Aoki95,Hao95,Sandvik99}. For
the $S=1$
SASLQHA, we show that $\alpha_c^{S=1}=0.040(5)$ is the quantum critical
point; we find that the thermodynamic properties at this point follow the
${\cal O}(3)$ QNL$\sigma$M predictions remarkably well. We compare three
quantities: the dimensionless ratio $S_Q/T\chi_s$ (defined below), the
temperature dependences of the correlation length $\xi$, and the uniform
susceptibility $\chi_u$. These quantities agree quantitatively with the
QNL$\sigma$M values without any adjustable parameter.


We have carried out quantum Monte Carlo simulations on large lattices
utilizing the loop cluster algorithm. The lattice size has been kept at
least 10 times larger than the calculated correlation length. The lengths
and Trotter numbers of the simulated lattices are chosen so as to minimize
any finite-size and lattice-spacing effects. Spin states are updated about
$2^{14}$ times to reach equilibrium and then measured $2^{15}$ times. The
same algorithm has been applied sucessfully in studying spin chains and
spin ladders. \cite{Greven96,Kim98,Kim99a} We compute the temperature (T)
and anisotropy ($\alpha$) dependence of the uniform susceptibility,
$\chi_u(\alpha,T)$; the correlation length, $\xi(\alpha,T)$; the staggered
susceptibility, $\chi_s(\alpha,T)$; and the static structure factor at the
antiferromagnetic wave vector $Q=(\pi,\pi)$, $S_Q(\alpha,T)$ for both the
$S=1/2$ and $S=1$ SASLQHA.

The structure of this paper is as follows:  Our Monte Carlo results for
the $S=1/2$ SASLQHA are presented in Sec.\ \ref{sec:half}.  The uniform
susceptibility and the correlation length data are shown and discussed
here. We present the $S=1$ SASLQHA Monte Carlo results in Sec.\
\ref{sec:one}. In Sec.\ \ref{sec:qc}, we determine and discuss the phase
diagram of the $S=1/2$ and $S=1$ SASLQHA using QNL$\sigma$M language,
especially focusing on the quantum critical behavior. In Sec.\
\ref{sec:discussion}, implications of this study in relation to
experiments and other related low-dimensional quantum magnets are
discussed.

\section{Results for S=1/2}
\label{sec:half}


In Fig.\ \ref{fig1}, we show the uniform susceptibility per spin for
$S=1/2$, as a function of $T$ for various $\alpha$ in semi-log scale. The
results for the $\alpha=1$ square lattice are taken from the recent Monte
Carlo study by Kim and Troyer. \cite{JKKim97} The crossover from 1D to 2D
behavior is clear in this figure. For small $\alpha$, the uniform
susceptibility follows that of a single chain, but begins to deviate
around $T/J \sim 5 \alpha$. For larger $\alpha$, $\chi_u$ lies
intermediate between those for a single chain and the square lattice, as
expected.

In the small $\alpha$ limit, the Monte Carlo results can be compared with
the results of experiments on weakly coupled spin chains. In particular,
Sr$_2$CuO$_3$ has a structure where $S=1/2$ spin chains lie along the
crystallographic $b$-direction. The interchain interaction along the
$a$-direction is frustrated, thus making this system quasi
two-dimensional. In the $bc$-plane, the interaction along the
$c$-direction is much smaller than that in the $b$-direction, therefore,
Sr$_2$CuO$_3$ is a very good realization of the $S=1/2$ SASLQHA in the
small $\alpha$ limit. Specifically, the crystallographic $b$-direction
corresponds to the y-direction in the notation of the Hamiltonian, Eq.\
(\ref{hamil}), while the $c$-direction corresponds to the x-direction.
Recently Motoyama and coworkers \cite{Motoyama96} measured the magnetic
susceptibility of Sr$_2$CuO$_3$. They extracted a value for the intrachain
exchange $J=2200(200)K$ by fitting the result to the theoretical
expression proposed by Eggert {\em et al.}. \cite{Eggert94} Motoyama {\em
et al.} also observed the logarithmic correction term predicted in the
theory. However, a sudden drop in the susceptibility near $20K$ was
observed, and this was attributed to the onset of three-dimensional (3D)
N\'eel order. In Fig.\ \ref{fig1}, the Sr$_2$CuO$_3$ experimental results
are shown as solid circles using $J=J_b=2200K$.  As one can see from the
figure, this drop can be naturally explained by a small but non-zero
interchain coupling. From a comparison with our Monte Carlo results, one
can estimate the interchain coupling in Sr$_2$CuO$_3$ as $J_c=J_x \approx
0.002J \approx 0.4$meV.

A couple of points need to be mentioned in estimating the interchain
coupling in Sr$_2$CuO$_3$. First, since the analysis of the susceptibility
data involves the subtraction of a Curie term due to impurities, subtle
effects may be difficult to interpret unambiguously. However, recent
nuclear magnetic resonance Knight shift measurements, in which the Knight
shift is proportional to the uniform susceptibility, show complete
agreement with the susceptibility data. \cite{NMR} Second, Schulz
\cite{Schulz96} and Wang \cite{Wang97} have predicted the staggered
magnetization of this system as a function of the interchain coupling,
$m_0=0.72(J_\perp / J)^{1/2}$ from their interchain mean-field
theory. \cite{Sandvik-log} With $\alpha=0.002$, we obtain $m_0= 0.032
\approx 0.064 \mu_B $, which agrees with the experimentally
\cite{Kojima97} determined value of $0.06(1) \mu_B$, giving credence to
our estimate of the interchain coupling in
Sr$_2$CuO$_3$.


The spin-spin correlation length is deduced from fits of the calculated
instantaneous spin-spin correlation function to the asymptotic
Ornstein-Zernike (OZ) form. Only large distance numerical data are
included in the fits to ensure that the asymptotic behavior is probed. As
discussed in our previous Monte Carlo studies of quantum spin systems,
\cite{Greven96,Kim98} the 1D OZ form works best at high temperatures,
while the 2D OZ form works better at low temperatures. However, for the
S=1 SASLQHA we observe a crossover to the 3D OZ form at low temperatures.
This crossover of the correlation function will be discussed in Sec.\
\ref{sec:discussion}.

The correlation lengths so-obtained for $S=1/2$ are plotted in Fig.\
\ref{fig2} on a logarithmic scale as a function of inverse temperature.
Therefore, linear behavior in this plot corresponds to an exponential
dependence on $T^{-1}$, and the slope corresponds to the spin stiffness.
At low enough temperatures, all data show linear behavior except for
$\alpha=0$, signaling that $\alpha=0$ is indeed a critical point. The
solid lines are the results of fits to the crossover form suggested 
by Castro Neto and Hone,
\cite{CastroNeto96} which interpolates between the low-$T$
expression calculated by Hasenfratz and Niedermayer \cite{Hasenfratz91}
for the (2+1) dimensional QNL$\sigma$M and $\xi \sim T^{-1}$ at high
temperature:
\begin{equation}
\xi =A \exp(2 \pi \rho_s (\alpha)/T) / \left[
1+\frac{1}{2} \frac{T}{2 \pi \rho_s (\alpha)}\right],
\label{eq:HN} 
\end{equation}
Two adjustable parameters are used in the fitting: A and $\rho_s$.  The
anisotropy dependence of $\rho_s(\alpha)$ is used to provide a crossover
temperature scale between quantum critical and renormalized classical
behavior in Sec.\ \ref{sec:qc}. We find that the $\rho_s(\alpha)$ values
extracted from fitting $\xi_x$ and $\xi_y$ agree within the errors, as
evidenced by the identical slopes in Fig.\ \ref{fig2}.

\section{Results for S=1}
\label{sec:one}

The uniform susceptibility data for $S=1$ are shown in Fig.\ \ref{fig3}.
The results for the square lattice ($\alpha=1$) are taken from recent QMC
work by Harada and coworkers. \cite{Harada98} As in the $S=1/2$ case, for
small $\alpha$ any deviation from the single chain data is only observed
at very low temperatures. As is evident in Fig.\ \ref{fig3}, one observes
distinctively different behavior for $\alpha=0.02$ compared with that for
$\alpha=0.05$. Specifically, $\chi_u (\alpha=0.02)$ drops to zero at low
temperatures, signaling the opening of a spin gap and concomitantly a spin
liquid ground state. Indeed, one can fit the data with the asymptotic form
$\chi_u(T) \sim \exp(-\Delta/T)$ at low temperatures. The result of the
fit is shown as the solid line for $\alpha=0.02$ in Fig.\ \ref{fig3}.  
The spin gap values from these fits are as follows:
$\Delta(\alpha=0.01)/J=0.32(1)$, $\Delta(\alpha=0.02)/J=0.25(1)$, and
$\Delta(\alpha=0.03)/J=0.17(1)$. Note that the $\alpha=0.01$ and
$\alpha=0.03$ data are not shown in Fig.\ \ref{fig3} for graphical
purposes.

In Fig.\ \ref{fig4}, the correlation length data in the y-direction are
plotted for $S=1$.  Since most correlation length data along the
x-direction are smaller than one lattice constant, they are not plotted.
Note the linear scale in this figure, unlike the logarithmic scale in
Fig.\ \ref{fig2} for $S=1/2$. This clearly shows the $1/T$ dependence of
the correlation length for $\alpha=0.04$, which thus identifies
$\alpha=0.04$ as a quantum critical point. For smaller values of $\alpha$
the correlation length saturates at low temperatures: $\xi_0
(\alpha=0.01)=7.0(2)$, $\xi_0 (\alpha=0.02)=10.8(5)$, $\xi_0
(\alpha=0.03)=19.0(5)$. For $\alpha > \alpha_c$, the correlation length
diverges exponentially in $1/T$. Therefore, one can fit the data to Eq.\
(\ref{eq:HN}) to obtain the spin stiffness $\rho_s(\alpha)$.

Some precautions are necessary in extracting the correlation length from
the correlation function via fits to the OZ form. The OZ form for the
correlation function in general is
\begin{equation}
C(r) \sim \frac{e^{-r/\xi}}{r^{\lambda}}, \;\;\;\; \lambda=\frac{(d-1)}{2},
\label{eq:OZ}
\end{equation}
where $d$ is the dimensionality. In quantum systems at low temperatures,
$d$ should be replaced by the effective dimensionality $(d+1)$. Since the
low-temperature and long-wavelength behavior of the $S=1$ SASLQHA with
non-zero $\alpha$ is mapped onto the (2+1) dimensional ${\cal O}(3)$
QNL$\sigma$M , the functional form of the correlation function at low
temperatures and long distances should be the OZ form with $\lambda=1$.  
Similarly, the 1D $S=1$ spin chain (SASLQHA with $\alpha=0$) has a
correlation function of the $\lambda=0.5$ OZ form. \cite{Kim98} We indeed
observe a crossover in the functional form of $C(r)$ from $\lambda=0.5$ to
$\lambda=1$ by increasing $\alpha$ from zero to a small but non-zero value
in the $S=1$ SASLQHA. In Fig.\ \ref{fig5}, the correlation function at the
low temperature $T/J=0.04$ is plotted to illustrate the difference in
$\lambda$. In order to show the subtle $\lambda$ dependence, $1/C(r)$ is
multiplied by the exponential factor, and only the $e^{-r/\xi}/\xi/C(r)
\sim r^{\lambda}$ part is shown as a function of $r/\xi$. The $\lambda=1$
behavior is apparent for $\alpha=0.025$, while $\lambda=0.5$ describes the
$\alpha=0$ data better. For comparison purposes, $C(r)$ for an $S=1/2$
chain ($\alpha=0$) is also plotted; in that case the $\lambda=0$ behavior
is quite clear.

One should note that this crossover is observed only in the QD phase,
since the finite correlation lengths in this phase allow the QMC technique
to probe the low temperature behavior. On the other hand, in phases with a
diverging correlation length, it is difficult to study the ground state
properties with the QMC method. Considering that the SASLQHA behaves
classically at high temperatures, it is possible that the $S=1/2$ data in
Fig.\ \ref{fig5} ($T/J=0.04$) have not reached low enough temperatures to
show the true ground state behavior of $\lambda=0.5$.  In deducing the
correlation length data shown in Fig.\ \ref{fig4}, the $\lambda=1$ form
was used for $\alpha \geq 0.02$ at low temperatures, while $\lambda=0.5$
was used otherwise.

\section{Quantum critical point}
\label{sec:qc}

The phase diagram of the $S=1$ SASLQHA is sketched in Fig.\ \ref{fig6},
where we use the energy gap $\Delta$ and spin stiffness $2 \pi \rho_s$,
obtained from the fits discussed in previous sections as the crossover
energy scales. We use the terminology of the ${\cal O}(3)$ QNL$\sigma$M to
identify the different phases in the figure. \cite{Chakravarty89} The
inverse anisotropy ratio $1/\alpha$ corresponds to the coupling constant
$g$. The $S=1/2$ phase diagram is almost identical to that of Ref.\
\onlinecite{Tworzydlo99}, and hence is not shown here.

At the quantum phase transition in (2+1) dimensions, $\rho_s$ and $\Delta$
obey the scaling laws
\begin{eqnarray}
\rho_s & \sim & (g_c - g)^{\nu} \label{eq:scaling1}\\
\Delta & \sim & (g-g_c)^{\nu},
\label{eq:scaling2}
\end{eqnarray}
with a critical exponent $\nu \approx 0.69$ for the 2DQHA.  
\cite{Troyer97} Since $\alpha^{-1}$ plays the role of the coupling
constant $g$, we plot $\Delta_0(\alpha-\alpha_c)^\nu$ as a shaded line in
Fig.\ \ref{fig6}. Note that $\Delta_0$ is fixed from the $S=1$ chain
value, so that there is no adjustable parameter. The power law scaling
fits the gap data quite well, confirming that the low-temperature
long-wavelength behavior of the $S=1$ SASLQHA is consistent with that of
the ${\cal O}(3)$ QNL$\sigma$M.

To illustrate that $\alpha_c=0.04$ is indeed the quantum critical point of
the $S=1$ SASLQHA, the dimensionless ratio $S_Q/T\chi_s$ is plotted in
Fig.\ \ref{fig7}(a). According to the QC scaling prediction for the ${\cal
O}(3)$ QNL$\sigma$M, \cite{Sokol94} this ratio should exhibit universal
behavior in the QC regime with the specific value $S_Q/T\chi_s=1.10(2)$.
As shown in Fig.\ \ref{fig7}(a), the $\alpha=0.04$ data for $S_Q/T\chi_s$
are constant around 1.1 at low temperatures, in quantitative agreement
with the QC scaling value. This ratio has been sucessfully used in
identifying quantum critical behavior in different systems, for example,
weakly coupled spin ladders.  \cite{Kim99a} We also show the correlation
length $\xi_y$ multiplied by temperature in Fig.\ \ref{fig7}(b), to
emphasize the $1/T$ dependence of $\xi_y$ at the quantum critical point,
$\alpha_c=0.04$, at low temperatures. Using the QNL$\sigma$M prediction
\cite{Chubukov94b} $\xi_y T=c/1.04$, we obtain the spin wave velocity of
$c_y/J=2.6$, which is slightly larger than the 1D value (2.49), but
smaller than the 2D value (3.067).

Another quantitative prediction from the ${\cal O}(3)$ QNL$\sigma$M is the
uniform susceptibility. In the QC regime the uniform susceptibility is
given as
\cite{Chubukov94b} 
\begin{equation}
\chi_u(T)=\Omega_1(\infty) \frac{1}{c^2} T,
\label{eq:unif-qc}
\end{equation}
where $\Omega_1(\infty)=0.26(1)$ is a universal constant. \cite{Troyer97}
Since the spin wave velocity is anisotropic in this case, one should
presumably use $c_xc_y$ instead of $c^2$ in Eq.\ (\ref{eq:unif-qc}).  The
inset of Fig.\ \ref{fig3} clearly shows the linear dependence of $\chi_u$
on temperature. The fitted value of the slope is 0.34(1). Since we do not
know the value of $c_x$, we use the fitted value of the slope and Eq.\
(\ref{eq:unif-qc}) to estimate $c_x$. If we take the spin wave velocity
value $c_y$ determined above, the spin wave velocity in the x-direction is
$c_x/J \approx 0.3$. This gives the ratio of the spin wave velocity
$c_x/c_y \approx 0.12$, which is very close to the value given by the
series expansion study \cite{Affleck94} and spin-wave theory:
\cite{Parola93} $c_x/c_y \approx 0.14$. One should note that the
anisotropy in the spin-wave velocity is enhanced from the mean field value
($c_x/c_y =\sqrt{\alpha}=0.2$) due to quantum fluctuations.

\section{Discussion}
\label{sec:discussion}

The most surprising result for the $S=1$ SASLQHA is the smallness of the
value $\alpha_c^{S=1} \approx 0.04 $, compared with the Haldane gap value
of the $S=1$ spin chain ($\Delta_H/J \approx 0.41$). One would naively
think that the interchain coupling should be comparable to the Haldane gap
to overcome the large energy gap and drive the system to long-range order.
Indeed, such heuristic reasoning works well in the case of coupled
even-legged spin ladders, where the value of the spin gap and the critical
interladder coupling values are $\sim 0.5J$ and $ \sim 0.3J$,
respectively, for an array of two-leg ladders. Corresponding values for an
array of four-leg ladders are $\sim 0.1J$ and $ \sim 0.07J$.

Recently there has been a number of studies on the nature of the ground
state of $S=1/2$ spin ladders. Specifically, much theoretical interest has
focused on the question of whether or not the ground state of the $S=1/2$
antiferromagnetic two-leg ladder is the same as that of the $S=1$ chain.
Since the $S=1$ chain is equivalent to the $S=1/2$ two-leg ladder with an
infinite ferromagnetic coupling, the above question can be rephrased as
whether the $S=1/2$ ladder goes through a phase transition when the
inter-chain coupling is varied from a positive (antiferromagnetic) value
to a negative (ferromagnetic) value.  White \cite{White96} has shown that
the antiferromagnetic ladder can be transformed continously to an $S=1$
chain by switching on an irrelevant next nearest neighbor coupling, thus
claiming the equivalence of the two. Kim and coworkers \cite{EKim99}
recently have studied a two-leg ladder with various model parameters by a
bosonization method and have provided evidence that the antiferromagnetic
ladder and $S=1$ chain belong to different universality classes, each
having distinct topological string order. Our results appear to show the
distinctively different nature of the ground state of the $S=1$ Haldane
chain and that of the antiferromagnetic ladder. However, more theoretical
and numerical studies are necessary to resolve this issue unambiguously.

\begin{figure}
\begin{center}
\epsfig{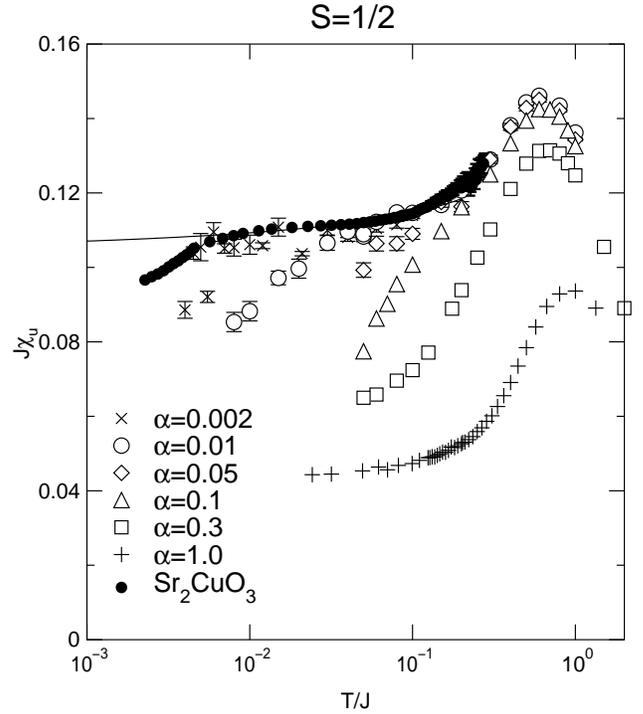}
\end{center}
\caption
{The uniform susceptibility per spin, $\chi_u$ for $S=1/2$, is shown as a
function of $T$ for various $\alpha$ in semi-logarithmic scale. $\alpha=1$
data are taken from Ref. 21. The Sr$_2$CuO$_3$ experimental results are
also shown; they are scaled to fit the Monte Carlo results and $J=2200K$
is used to scale temperature. The solid line is a plot of the field theory
result from Ref. 23. }
\label{fig1}
\end{figure}

\begin{figure}
\begin{center}
\epsfig{file=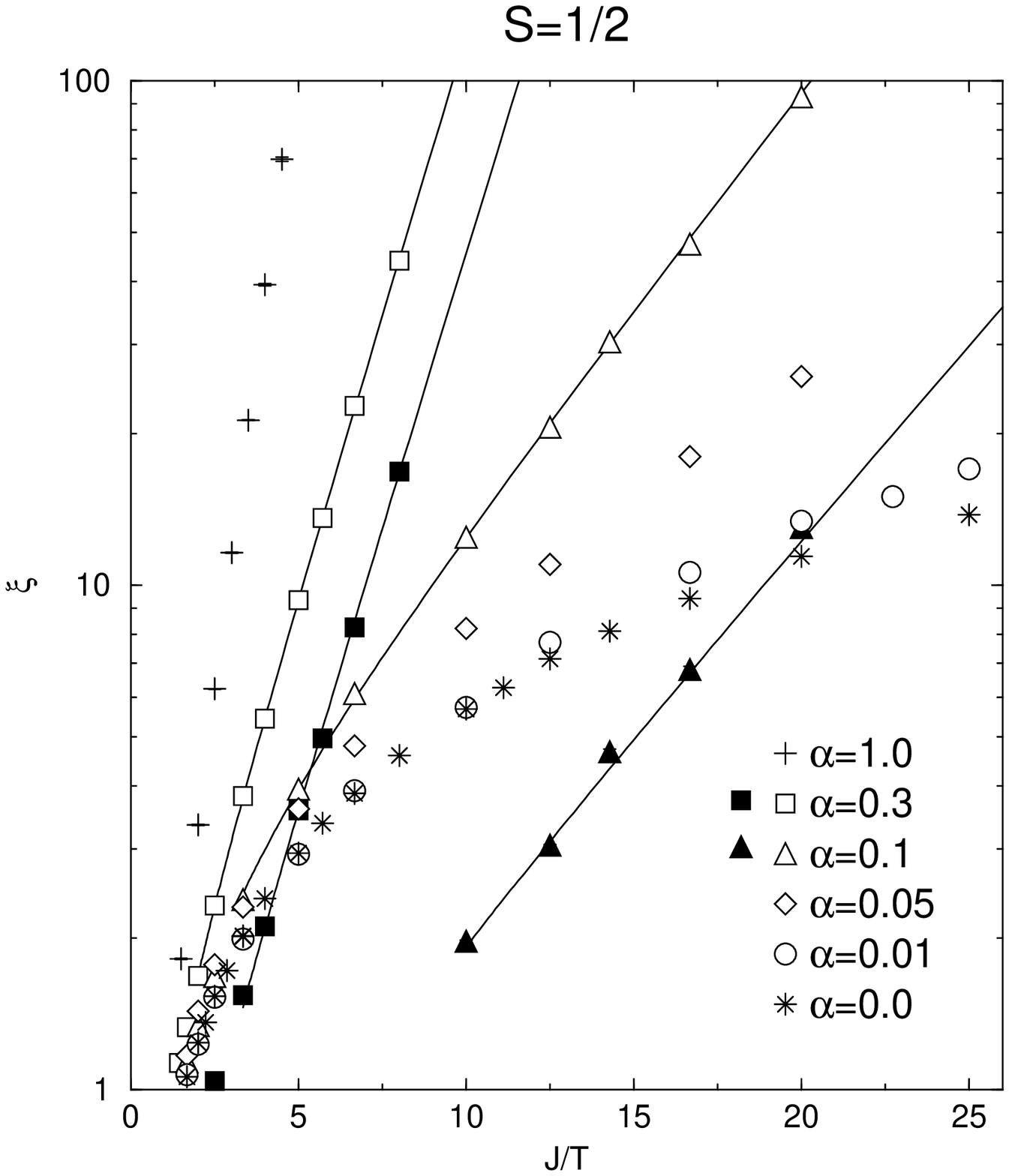,width=\figwidth}
\end{center}
\caption
{The correlation length for $S=1/2$ as a function of $J/T$ for various
values of $\alpha$ in a logarithmic scale.  Filled symbols denote the
correlation length in the x-direction ($\xi_x$), while open symbols are
$\xi_y$ data. For small $\alpha$ ($\alpha \leq 0.05$), the $\xi_x$'s are
smaller than one lattice constant and are not shown in the figure. Solid
lines are fits to Eq.\ (\ref{eq:HN}), showing the exponential dependence of
$\xi$ on $T^{-1}$. }
\label{fig2}
\end{figure}

\begin{figure}
\begin{center}
\epsfig{file=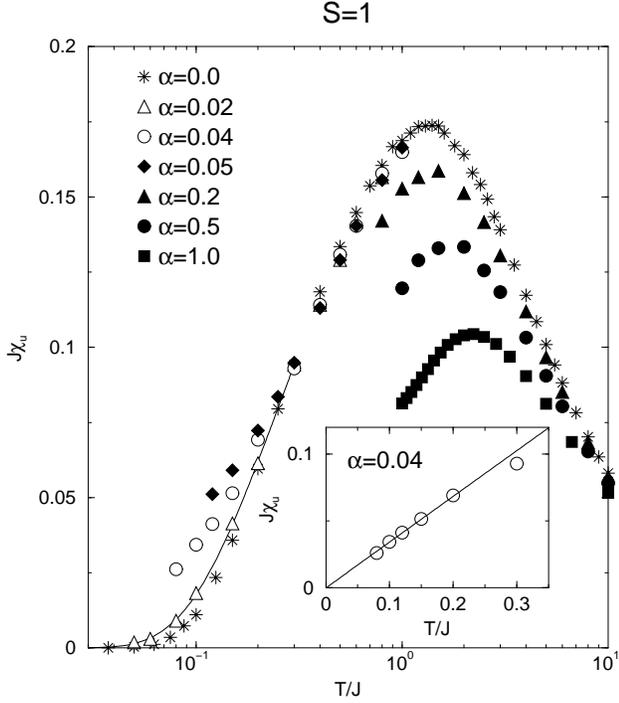,width=\figwidth}
\end{center}
\caption
{The uniform susceptibility per spin, $\chi_u$ for $S=1$, shown as a
function of $T$ for various values of $\alpha$ on a semi-logarithmic
scale. The $\alpha=1$ data are taken from Ref. 28. The solid line is the
result of a fit to the asymptotic form $\chi_u \sim \exp(-\Delta/T)$.
Inset: The $\alpha=0.04$ data on a linear scale. The solid line is
$J\chi_u=0.34T/J$.}
\label{fig3}  
\end{figure}

\begin{figure}
\begin{center}
\epsfig{file=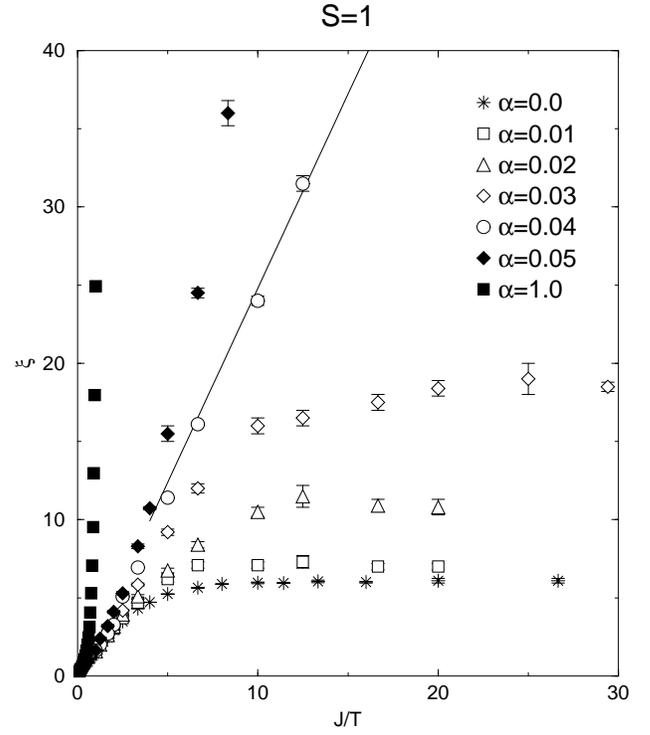,width=\figwidth}
\end{center}
\caption
{The correlation length along the y-direction, $\xi_y$, for $S=1$ as a
function of $J/T$ for various values of $\alpha$. Note that this plot is
on a linear scale, so that the straight line through the $\alpha=0.04$
data signifies the $1/T$ dependence of $\xi$.  Filled symbols show
exponentially diverging correlation lengths, while open symbols show the
saturation of the correlation length at low temperature for a QD ground
state.  }
\label{fig4}
\end{figure}

\begin{figure}
\begin{center}
\epsfig{file=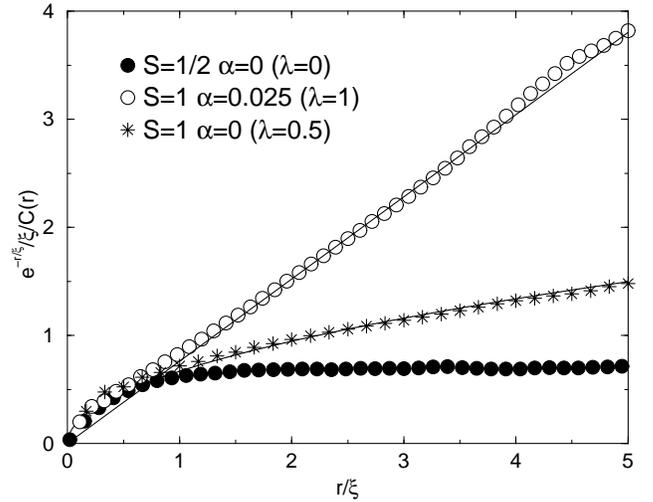,width=\figwidth}
\end{center}
\caption
{Inverse correlation function multiplied by the exponential factor
$e^{-r/\xi}/\xi/C(r)$ plotted as a function of distance $r/\xi$. The solid
lines are the results of fits to the form $\sim r^\lambda$ with $\lambda$
fixed at the given values. All data are taken at $T/J=0.04$.}
\label{fig5}
\end{figure}

\begin{figure}
\begin{center}
\epsfig{file=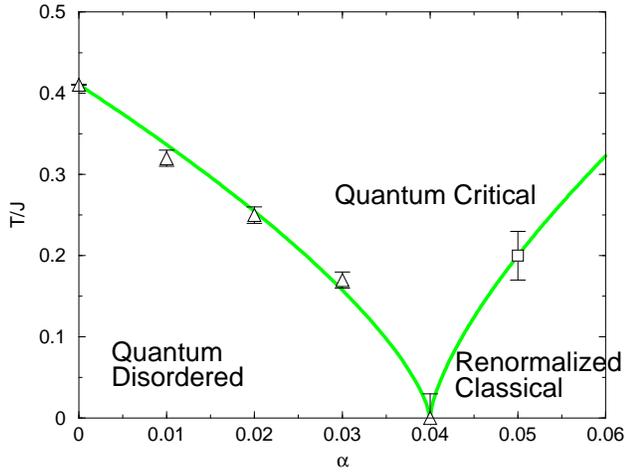,width=\figwidth}
\end{center}
\caption
{Phase diagram of the $S=1$ SASLQHA. The open squares are $2\pi \rho_s/J$
obtained by fitting the correlation length to the asymptotic form Eq.\
(\ref{eq:HN}), and the open triangles are the energy gap $\Delta/J$ obtained
by fitting the uniform susceptibility to the form $\sim \exp(-\Delta/T)$.
The shaded lines are the power laws, Eq.\ (\ref{eq:scaling1}) and Eq.\
(\ref{eq:scaling2}).
}
\label{fig6}
\end{figure}

\begin{figure}
\begin{center}
\epsfig{file=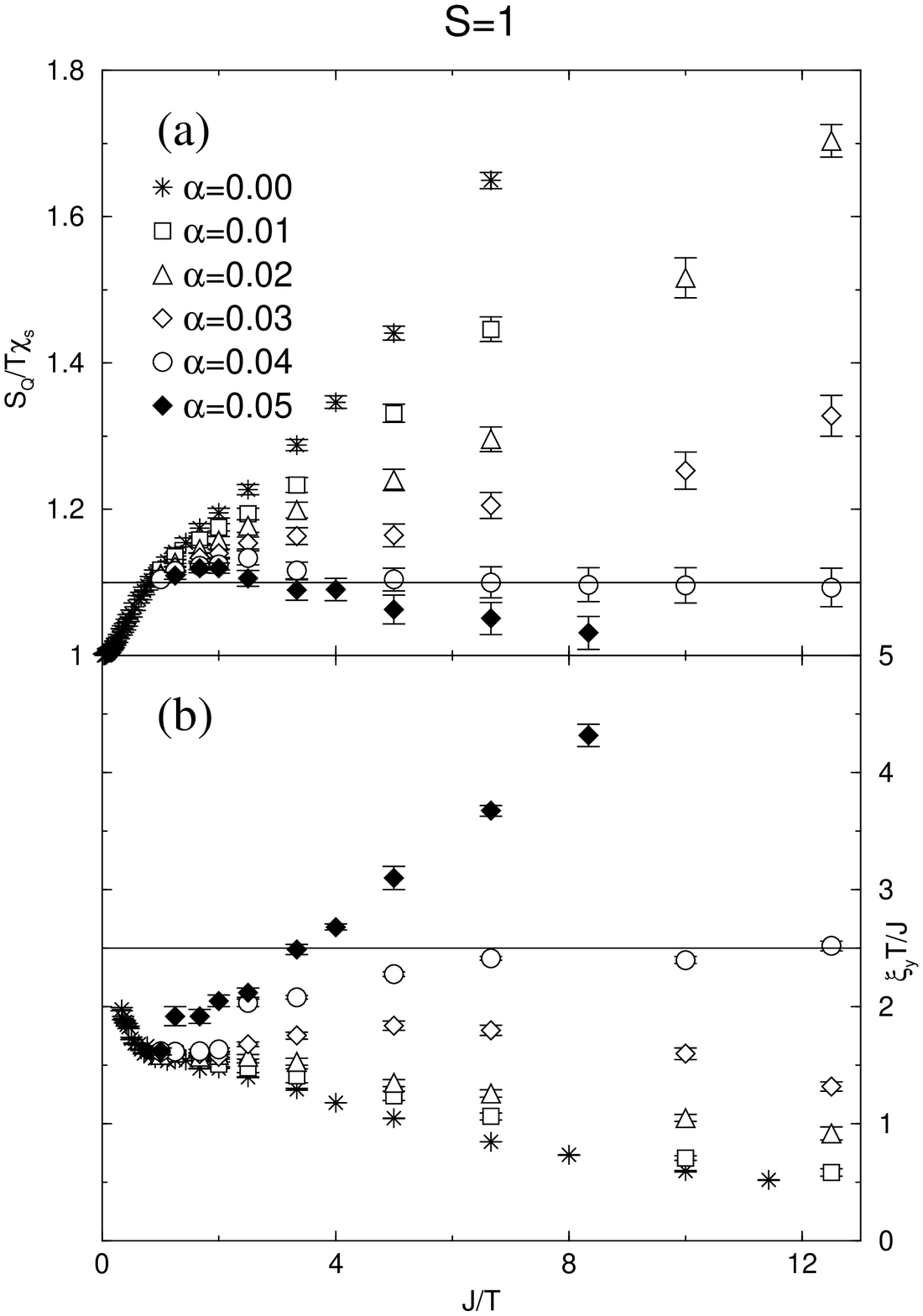,width=\figwidth}
\end{center}
\caption
{(a) Dimensionless ratio $S_Q/T\chi_s$ as a function of inverse
temperature for various values of $\alpha$. The solid line is the
prediction from the ${\cal O}(3)$ QNL$\sigma$M (b) Correlation length
multiplied by temperature as a function of inverse temperature. The solid
line corresponds to the value $c_{1D}=2.49$. }
\label{fig7}  
\end{figure}

Although both the SASLQHA and spin ladders show crossovers between 1D and
2D behavior, the intrinsic difference between the two approaches should be
noted. At high temperatures, the SASLQHA behaves as an array of decoupled
chains. The 2D behavior is only observed at low temperatures, as shown in
Fig.\ \ref{fig1} and Fig.\ \ref{fig5}. However, the spin ladders at high
temperatures show essentially 2D physics, since the spin-spin correlation
length is shorter than the width of the spin ladder. Therefore, the
crossover for the spin ladders (2D $\rightarrow$ 1D as T $\rightarrow$ 0)
is complementary to that of the SASLQHA.

In Sec.\ \ref{sec:half}, we estimated the exchange interaction between
chains in Sr$_2$CuO$_3$ as $\sim 0.4$ meV. Greven and Birgeneau have
considered a similar exchange interaction for SrCu$_2$O$_3$.
\cite{Greven98} Namely, they have noted that the the effective coupling
between copper spins in different ladder planes is mediated by Sr$^{2+}$
ions and they have argued that this interladder exchange should be about
10 meV. This value is also used to describe the behavior of the three leg
ladder compound Sr$_2$Cu$_3$O$_5$. \cite{Thurber99} This represents quite
a large discrepancy between values estimated for seemingly similar
superexchange interactions. However, one should observe that the geometry
of the inter-plane coupling in the spin ladder compound SrCu$_2$O$_3$ and
that of the inter-chain coupling in the spin chain compound Sr$_2$CuO$_3$
are quite different. Specifically, in Sr$_2$CuO$_3$ , all
Cu$^{2+}$--Sr$^{2+}$--Cu$^{2+}$ bond angles are close to 90$^\circ$, while
there exist linear Cu$^{2+}$--Sr$^{2+}$--Cu$^{2+}$ bonds in SrCu$_2$O$_3$
and Sr$_2$Cu$_3$O$_5$. If one assumes that the coupling between these
Cu$^{2+}$ ions is primarily due to the superexchange mediated by Sr$^{2+}$
ions, one can infer that the effective inter-plane coupling in
SrCu$_2$O$_3$ is mostly due to the 180$^\circ$
Cu$^{2+}$--Sr$^{2+}$--Cu$^{2+}$ superexchange interaction. This leads us
to speculate that the small value ($\sim 0.4$ meV) of the interchain
coupling in Sr$_2$CuO$_3$ is due to the absence of linear
Cu$^{2+}$--Sr$^{2+}$--Cu$^{2+}$ superexchange paths.

This realization is important in understanding the spin fluctuations in
superconducting YBa$_2$Cu$_3$O$_{6+\delta}$, since the bilayer coupling
path in YBa$_2$Cu$_3$O$_{6+\delta}$ is similar to the inter-plane coupling
path in the spin ladder compound SrCu$_2$O$_3$. \cite{Reznik96} Although
the detailed nature of the hopping integral between bilayers is not known,
Andersen {\it et al.} have shown that the hopping mediated by the yttrium
ion contributes significantly to the bilayer exchange coupling
\cite{Andersen95}. Obviously, more theoretical calculations of the bilayer
exchange interaction are needed.

In summary, we have studied the $S=1/2$ and $S=1$ SASLQHA with the quantum
Monte Carlo method. By varying the anisotropy $\alpha$ from 0 to 1, we go
continuously from the one-dimensional chain to the two-dimensional
isotropic square lattice.  The temperature dependence of the uniform
susceptibility and the spin-spin correlation length are presented for
various values of $\alpha$. For $S=1$, we show that there exists a quantum
critical point at $\alpha^{S=1}_c=0.040(5)$, in agreement with other
analytic predictions. The power law behavior of the spin gap is also
confirmed. In addition, universal quantities predicted from the
QNL$\sigma$M, such as $S_Q/T\chi_s \approx 1.1$ and $\Omega_1(\infty)  
\approx 0.26$, agree with our results at $\alpha=0.04$, without any
adjustable parameter.  The $S=1/2$ results are consistent with
$\alpha^{S=1/2}_c=0$, as discussed by a number of previous theoretical
studies. We also estimate the interchain coupling in the $S=1/2$ compound
Sr$_2$CuO$_3$ by comparing the measured uniform susceptibility data with
our QMC results.

\acknowledgements{We would like to thank M. Greven for invaluable
discussions. This work was supported by the National Science
Foundation--Low Temperature Physics Programs under award number DMR
97--15315.}

\begin{thebibliography}{10}

\bibitem{Bethe31}
H.~A. Bethe, Z. Phys. {\bf 71},  205  (1931).

\bibitem{Haldane83a}
F.~D.~M. Haldane, Phys. Lett. {\bf 93A},  464  (1983).

\bibitem{Chakravarty89}
S. Chakravarty, B.~I. Halperin, and D.~R. Nelson, Phys. Rev. B {\bf 39},  2344
  (1989).

\bibitem{Dagotto96}
E. Dagotto and T.~M. Rice, Science {\bf 271},  618  (1996).

\bibitem{CastroNeto96}
A.~H. {Castro Neto} and D. Hone, Phys. Rev. Lett. {\bf 76},  2165  (1996).

\bibitem{vanDuin98}
C.~N.~A. {van Duin} and J. Zaanen, Phys. Rev. Lett. {\bf 80},  1513  (1998).

\bibitem{Parola93}
A. Parola, S. Sorella, and Q.~F. Zhong, Phys. Rev. Lett. {\bf 71},  4393
  (1993).

\bibitem{Sakai89}
T. Sakai and M. Takahashi, J. Phys. Soc. Jpn {\bf 58},  3131  (1989).

\bibitem{Azzouz93a}
M. Azzouz and B. Dou\c{c}ot, Phys. Rev. B {\bf 47},  8660  (1993).

\bibitem{Azzouz93b}
M. Azzouz, Phys. Rev. B {\bf 48},  6136  (1993).

\bibitem{Affleck94}
I. Affleck, M.~P. Gelfand, and R.~R.~P. Singh, J. Phys. A {\bf 27},  7313
  (1994).

\bibitem{Affleck96}
I. Affleck and B.~I. Halperin, J. Phys. A {\bf 29},  2627  (1996).

\bibitem{Miyazaki95}
T. Miyazaki, D. Yoshioka, and M. Ogata, Phys. Rev. B {\bf 51},  2966  (1995).

\bibitem{Aoki95}
T. Aoki, J. Phys. Soc. Jpn {\bf 64},  605  (1995).

\bibitem{Hao95}
B. Hao and C. Gong, Phys. Rev. B {\bf 52},  299  (1995).

\bibitem{Sandvik99}
A.~W. Sandvik, Phys. Rev. Lett. {\bf 83},  3069  (1999).

\bibitem{Greven96}
M. Greven, R.~J. Birgeneau, and U.-J. Wiese, Phys. Rev. Lett. {\bf 77},  1865
  (1996).

\bibitem{Kim98}
Y.~J. Kim, M. Greven, U.-J. Wiese, and R.~J. Birgeneau, Eur. Phys. J. B {\bf
  4},  291  (1998).

\bibitem{Kim99a}
Y.~J. Kim, R.~J. Birgeneau, M.~A. Kastner, Y.~S. Lee, Y. Endoh, G. Shirane, and
  K. Yamada, Phys. Rev. B {\bf 60},  3294  (1999).

\bibitem{JKKim97}
J.~K. Kim and M. Troyer, Phys. Rev. Lett. {\bf 80},  2705  (1998).

\bibitem{Motoyama96}
N. Motoyama, H. Esaki, , and S. Uchida, Phys. Rev. Lett. {\bf 76},  3212
  (1996).

\bibitem{Eggert94}
S. Eggert, I. Affleck, and M. Takahashi, Phys. Rev. Lett. {\bf 73},  332
  (1994).

\bibitem{NMR}
K.~R. Thurber, private communications  .

\bibitem{Schulz96}
H. Schulz, Phys. Rev. Lett. {\bf 77},  1790  (1996).

\bibitem{Wang97}
Z. Wang, Phys. Rev. Lett. {\bf 78},  126  (1997).

\bibitem{Sandvik-log}
In Ref.\ \onlinecite{Sandvik99}, Sandvik has obtained logarithmic
correction to the simple mean-field behavior.

\bibitem{Kojima97}
K.~M. Kojima, Y. Fudamoto, M. Larkin, G.~M. Luke, J. Merrin, B. Nachumi, Y.~J.
  Uemura, N. Motoyama, H. Eisaki, S. Uchida, K. Yamada, Y. Endoh, S. Hosoya,
  B.~J. Sternlieb, and G. Shirane, Phys. Rev. Lett. {\bf 78},  1787  (1997).

\bibitem{Hasenfratz91}
P. Hasenfratz and F. Niedermayer, Phys. Lett. B {\bf 268},  231  (1991).

\bibitem{Harada98}
K. Harada, M. Troyer, and N. Kawashima, J. Phys. Soc. Jpn {\bf 67},  1130
  (1998).

\bibitem{Tworzydlo99}
J. Tworzydlo, Y. Osman, C.~N.~A. van Duin, and J. Zaanen, Phys. Rev. B {\bf
  59},  115  (1999).

\bibitem{Troyer97}
M. Troyer, M. Imada, and K. Ueda, J. Phys. Soc. Jpn {\bf 66},  2957  (1997).

\bibitem{Sokol94}
A. Sokol, R.~L. Glenister, and R.~R.~P. Singh, Phys. Rev. Lett. {\bf 72},  1549
   (1994).

\bibitem{Chubukov94b}
A.~V. Chubukov, S. Sachdev, and J. Ye, Phys. Rev. B {\bf 49},  11919  (1994).

\bibitem{White96}
S.~R. White, Phys. Rev. B {\bf 53},  52  (1996).

\bibitem{EKim99}
E.~H. Kim, G. F{\'a}th, J. S{\'o}lyom, and D.~J. Scalapino, cond-mat/9910023  .

\bibitem{Greven98}
M. Greven and R.~J. Birgeneau, Phys. Rev. Lett. {\bf 81},  1945  (1998).

\bibitem{Thurber99}
K.~R. Thurber, T. Imai, T. Saitoh, M. Azuma, M. Takano, and F.~C. Chou, Phys.
  Rev. Lett. {\bf 84},  558  (2000).

\bibitem{Reznik96}
D. Reznik, P. Bourges, H.~F. Fong, L. Regnault, J. Bossy, C. Vettier, D.~L.
  Milius, I.~A. Aksay, and B. Keimer, Phys. Rev. B {\bf 53},  R14741  (1996).

\bibitem{Andersen95}
O.~K. Andersen, A.~I. Liechtenstein, O. Jepsen, and F. Paulsen, J. Phys. Chem.
  Solids {\bf 56},  1573  (1995).

\end{thebibliography}
\end{document}